\documentstyle{mn}

\def\ltord{\hbox{$\;\raise.4ex\hbox{$<$}\kern-.75em\lower.7ex\hbox{$\sim$}
                       \;$}}

\title[Are Q-stars a serious threat for stellar-mass black hole candidates?]
{Are Q-stars a serious threat for stellar-mass black hole candidates?}
\author[J.C.~Miller, T.~Shahbaz and L.A.~Nolan]
{J.C.~Miller,$^{1,2}$ T.~Shahbaz$^{2}$ and L.A.~Nolan$^{2}$ \\
$^{1}$SISSA, via Beirut 2-4, 34013 Trieste, Italy \\
$^{2}$Nuclear and Astrophysics Laboratory, University of Oxford, 
Keble Road, Oxford OX1 3RH }

\begin{document}

\maketitle

\begin{abstract}
\noindent
We examine the status of the threat posed to stellar-mass black hole
candidates by the possible existence of Q-stars (compact objects with an
exotic equation of state which might have masses well above the
normally-accepted maximum for standard neutron stars). We point out that
Q-stars could be extremely compact (with radii less than 1.5 times the 
corresponding Schwarzschild radius) making it quite difficult to determine
observationally that a given object is a black hole rather than a Q-star,
unless there is direct evidence for the absence of a solid surface. On the
other hand, in order for a Q-star to have a mass as high as that inferred
for the widely-favoured black hole candidate V404 Cygni, it would be
necessary for the Q-matter equation of state to apply already at densities
an order of magnitude below that of nuclear matter and this might well be
considered implausible on physical grounds. We also describe how rotation
affects the situation and discuss the prospects for determining
observationally that black hole candidates are not Q-stars.  
\end{abstract}

\section{Introduction}

The search for an unequivocal demonstration of the existence of
stellar-mass black holes has focussed on X-ray emitting binary systems
consisting of an ordinary star together with a compact object. If the mass
of the compact object (as determined from kinematical measurements)
appears to be greater than the maximum possible for a neutron star
$M_{max}$, then the object has been acknowledged as a black hole
candidate. The value of $M_{max}$ is still not reliably known, because of
uncertainties in the equation of state of matter at high densities, but it
has been widely believed that the limit of $3.2\,M_\odot$ derived by
Rhoades \& Ruffini (1974) (together with a possible 25\% upward
correction for rotation) gives a secure upper bound. 

The best current stellar-mass black hole candidates are in soft X-ray
transients (SXTs), a sub-class of the low mass X-ray binaries (see Charles
1996).  During quiescence, the accretion disc in these systems becomes
extremely faint and it is then possible to carry out detailed photometry
and spectroscopy of the optical companion, which allows the mass of the
compact object to be directly determined (van Paradijs \& McClintock
1995). Fig.~1 shows the presently-known masses of neutron stars and
black-hole candidates. The ``neutron star'' masses all lie within a small
range of $1.4\, M_\odot$, whereas the black-hole candidates seem to form a
distinct grouping around $\sim 10\, M_\odot$. The most convincing of the
black hole candidates are V404 Cyg (Shahbaz {\it et al.} 1994b) and Nova
Sco (Orosz \& Bailyn 1996) with masses of $12\, M_\odot$ and $7\, M_\odot$
respectively, well above the Rhoades/Ruffini limit. 

Rhoades and Ruffini derived their result in response to the problem caused
by different high-density equations of state leading to widely different
values for $M_{max}$, the idea being to derive a firm {\it upper limit}
for non-rotating models on the basis only of knowledge which could be
considered as completely secure.  However, some of the assumptions made
are, in fact, distinctly questionable (see Hartle 1978, Friedman \& Ipser
1987). In particular: (i) it was assumed that the equation of state can be
taken as accurately known for densities up to a fiducial value $\rho_0$
which they took as $4.6 \times 10^{14}\,{\rm g}\,{\rm cm}^{-3}$; (ii) they
imposed a causality condition which would only be appropriate for a
non-dispersive medium. If a more conservative value is taken for $\rho_0$
($10^{14}\,{\rm g}\,{\rm cm}^{-3}$), the causality condition is dropped
and allowance is made for rotation, the upper bound for $M_{max}$ obtained
in this way goes up to $14.3\,M_\odot$ (Friedman \& Ipser 1987) which is
no longer useful in considering black hole candidates such as V404 Cyg.
However, standard realistic neutron star equations of state do, in
practice, give masses satisfying the original condition $M \ltord
3.2\,M_\odot$ even with rotation. (It should be noted that all of this
discussion of the mass limit is being made within the context of general
relativity which is taken to provide a correct description of gravity
within this strong-field regime. We work within this context throughout
the present paper. However, while general relativity is well-accepted as
our theory of gravity, we recall that it has not yet received convincing
experimental verification away from the weak-field limit and so one should
not overlook the possibility that this might {\it not} be the correct
description.)

How much can the standard neutron star equations of state be trusted for
matter at high densities?  Is it correct that neutron star matter consists
of a mixture of neutrons, protons, electrons, mesons, hyperons, etc., and
is held together just by its own self-gravity? Quantum chromodynamics
(QCD) contains the idea of {\it confinement} by the strong force which is
normally thought of in terms of quarks being confined within nucleons. It
was in connection with this that the {\it strange star} model was
introduced (Witten 1984; Haensel, Zdunik \& Schaeffer 1986; Alcock, Farhi
\& Olinto 1986), resting on the hypothesis that strange quark matter might
be the absolute ground state of baryonic matter even at zero pressure.
Strange stars would be essentially single giant nucleons with baryon
number $A \sim 10^{57}$ and with the confined quarks being free to move
within the false vacuum which extends throughout the interior. They could
have masses and radii similar to those of standard neutron stars and have
been advocated as a viable alternative model for pulsars although there is
a problem over explaining glitches.

Strange stars are ``safe'' as far as the mass limit is concerned. Although
their equation of state is very different from that for standard neutron
star matter, the maximum mass is within the standard range (it is $\sim
2.0\,M_\odot$ for non-rotating models with some variation depending on
uncertain parameter values). However, some effective field theories of the
strong force allow for it not only to confine quarks in the normal way but
also to confine {\it nucleons} (neutrons and protons) at densities well
below that of nuclear matter ($\rho_{nm} \sim 2.7 \times 10^{14}\,{\rm
g}\,{\rm cm}^{-3}$), giving an equation of state different from the
standard one at densities below the values normally taken for $\rho_0$.
Models based on this idea were introduced by Bahcall, Lynn \& Selipsky
(1989a,b, 1990), who named them {\it Q-stars} (although note that the
``Q'' here does not stand for ``quark'' but for a conserved particle
number). These are {\it not} safe for the mass limit and might, in
principle, have {\it very} high masses up to more than $100\,M_\odot$. 

Even if one rules out Q-stars as models for pulsars (they have similar
difficulties in this respect as for strange stars) there still remains the
possibility that they could be an alternative model for the more massive
objects in black-hole-candidate systems. In connection with this, there
are a number of questions which immediately present themselves. For a
given mass, how {\it small} could a Q-star be?  How {\it physically
reasonable} are the versions of the Q-star equation of state which would
allow masses as high as that of the compact object in V404 Cyg?  What
happens when {\it rotation} of the object is considered?  These issues
are considered in the next Section.

\section{Properties of Q-star models}

For our calculations, we have used the simplest form of the Q-star
equation of state:
$$
\rho - 3p - 4U_0 + \alpha_v ( \rho - p - 2U_0 )^{3/2} = 0
$$
where $\rho$ is the density, $p$ is the pressure, $U_0$ is the energy
density of the confining scalar field and $\alpha_v$ measures the strength
of the repulsive interaction between nucleons.  This represents chiral
Q-matter (for which the particles have zero mass within the false vacuum)
but the results obtained are only marginally different for non-chiral
Q-matter. It is convenient to introduce the parameter $\zeta$ $( =
\alpha_v U_0^{1/2} \pi / \sqrt{3})$ since, when this is fixed, all results
scale with $U_0$. 

Fig.~2 shows representative mass/radius curves for Q-stars and for a
standard neutron star equation of state (model C from the collection of
Arnett \& Bowers 1977) and it is clear that the two curves are quite
different. The lowest-mass Q-star models, for any given $\zeta$, have an
almost constant value of the density, giving $M \propto R^3$, but with
increasing mass, the density profile becomes progressively more peaked
towards the centre. As the central density is increased further, a maximum
mass is eventually reached and this marks the end of models which are
stable under radial perturbations. The maximum-mass model is the most
compact stable one ({\it i.e.} $R/M$ is a minimum) and therefore the
closest in size to a black hole of the same mass. In Fig.~3, we show
$(R/M)_{min}$ plotted as a function of $\zeta$ for Q-stars together with
corresponding values for a black hole and for the representative neutron
star equation of state. (We are using the standard geometrical units of
general relativity with $c = G = 1$.) It can be seen that {\it Q-stars can
be very compact} with $R/M < 3$ (although note that very large values of
$\zeta$ are not physically reasonable.) A stable, non-rotating
$12\,M_\odot$ Q-star might have a radius as small as $\sim 52\,$km as
compared with $\sim 36\,$km for a non-rotating black hole of the same
mass. Having $R/M < 3$ means that the surface lies inside the circular
photon orbit, the place where centrifugal force becomes attractive rather
than repulsive (see Abramowicz \& Prasanna 1990), but the models do not
seem to be quite compact enough to allow for some more exotic relativistic
effects such as resonance of axial gravitational wave modes (Chandrasekhar
\& Ferrari 1991) or production of an internal ergoregion when the object
is set into rapid rotation (see Butterworth \& Ipser 1976), which would
lead to an instability. 

The maximum mass which a Q-star could have is related to the threshold
minimum density, $\rho_{min}$, above which the Q-matter equation of state
is taken to apply. (This is the density at which the pressure goes to
zero.) The relation is shown in Fig.~4 for the limiting cases $\zeta = 0$
and $\zeta \to \infty$ and it can be seen that varying $\zeta$ over the
entire range in between makes only a small difference.  In order to have a
non-rotating Q-star with as high a mass as that inferred for the compact
object in V404 Cyg (marked with the horizontal dashed line), it would be
necessary for the threshold density to be below $\rho_{nm}$ (marked with
the vertical dashed line) by about a factor of 10 or more. Rotation makes
little difference to this. Even if one admits the idea of Q-matter in
principle for some range of densities around $\rho_{nm}$, such a very low
value of $\rho_{min}$ might seem implausible. 

Compact objects accreting matter in binary systems will be spun up by the
angular momentum of the accreted material and so it is important to ask
what effect this would have for the comparison between the black hole and
Q-star pictures. Fig.~5 shows the locations of the equatorial radii of a
black hole, an extreme Q-star ($\zeta \to \infty$, maximum compactness)
and a representative neutron star, as functions of $a/M$ where $a$ is the
angular momentum per unit mass and all quantities are again in geometrical
units. All of the radii are calculated in the same coordinate-independent
way, by taking the proper distance round the circumference of the circle
concerned and dividing by $2\pi$. Doing this, the equator of the black
hole event horizon is always at $r = 2M$ irrespective of the value of
$a/M$. (We emphasize that this is in contrast with the situation for the
commonly-used measure in Boyer-Lindquist coordinates for which the equator
is at $r = 2M$ when $a = 0$ but $r \to M$ as $a \to M$. Other
frequently-quoted quantities such as observed rotational frequencies and
binding energies are usually calculated in an invariant way.) The formulae
used for the Q-star and neutron star models are calculated within the
slow-rotation approximation, correct to second order in the rotational
velocity $\Omega$ and comparing rotating and non-rotating models with
equal central density. Also shown in Fig.~5 is the location of the
marginally stable orbit (corresponding to the inner edge of a Keplerian
accretion disc) which, for a given $a/M$, is the same for all of the
objects within the slow rotation approximation. While bearing in mind that
results obtained with the slow-rotation approximation should only be
regarded as indicative for the higher values of $a/M$, several interesting
conclusions can be drawn from this figure. As $a/M$ is increased, the
$r/M$ of the marginally stable orbit decreases and eventually it reaches
the equator of the neutron star which expands out to meet it. For the
extreme Q-star, $r/M$ of the equator actually {\it decreases} slightly and
the marginally stable orbit would reach it only for larger $a/M$. 

\section{Discussion and Conclusions} 

How can one be sure that a high-mass compact object such as that in V404
Cygni is a black hole and not a Q-star? Any evidence for the existence of
a solid surface ({\it e.g.} by observing a Type I X-ray burst) would, of
course, immediately rule out the possibility of the object being a black
hole but in the absence of such evidence, what can be done? Narayan,
McClintock \& Yi (1996) have proposed a model for some SXTs in quiescence
which is in agreement with available observational data and {\it depends}
on the compact object in question being a black hole. However, this is
still an indirect argument. What one would really like is to have direct
observational evidence of accreting material at radii smaller than would
be possible with a Q-star. With satellite experiments such as the Rossi
X-ray Timing Experiment (RXTE) and the Unconventional Stellar Aspect
experiment (USA) being capable of time resolution down to the order of
$1\,\mu{\rm sec}$, there is a possibility of detecting the location of the
inner edge of the accretion disc (if, indeed, there {\it is} a
well-defined inner edge) by seeing a corresponding frequency cut-off in
the power spectrum, by seeing dying pulse trains from bright features as
they reach the inner edge and fall in (Stoeger 1980) or from analysis of
quasi-periodic oscillations (Miller, Lamb \& Psaltis 1997; see Zhang {\it
et al.} 1997 for discussion of a possible detection of the inner edge of
the disc by this means for binaries with compact components in the neutron
star mass range). For non-rotating compact objects, the marginally stable
orbit is at $r = 6M$, much larger than the radius of the most interesting
Q-star models, but with increasing rotation of the compact object, it
moves inwards and if the compact object were a Q-star, it would eventually
meet the surface. If evidence were to be found for an accretion flow
extending further inwards than would be possible with a Q-star, this would
then point very strongly towards the compact object being a black
hole, since the Q-star model probably presents the last {\it realistic} 
possibility for avoiding that conclusion within the context of general 
relativity.

\section*{Acknowledgements}

Financial support from the Italian Ministero dell'Universit\`a e della
Ricerca Scientifica e Tecnologica and from PPARC are gratefully
acknowledged.

\section*{Figure captions}

\noindent
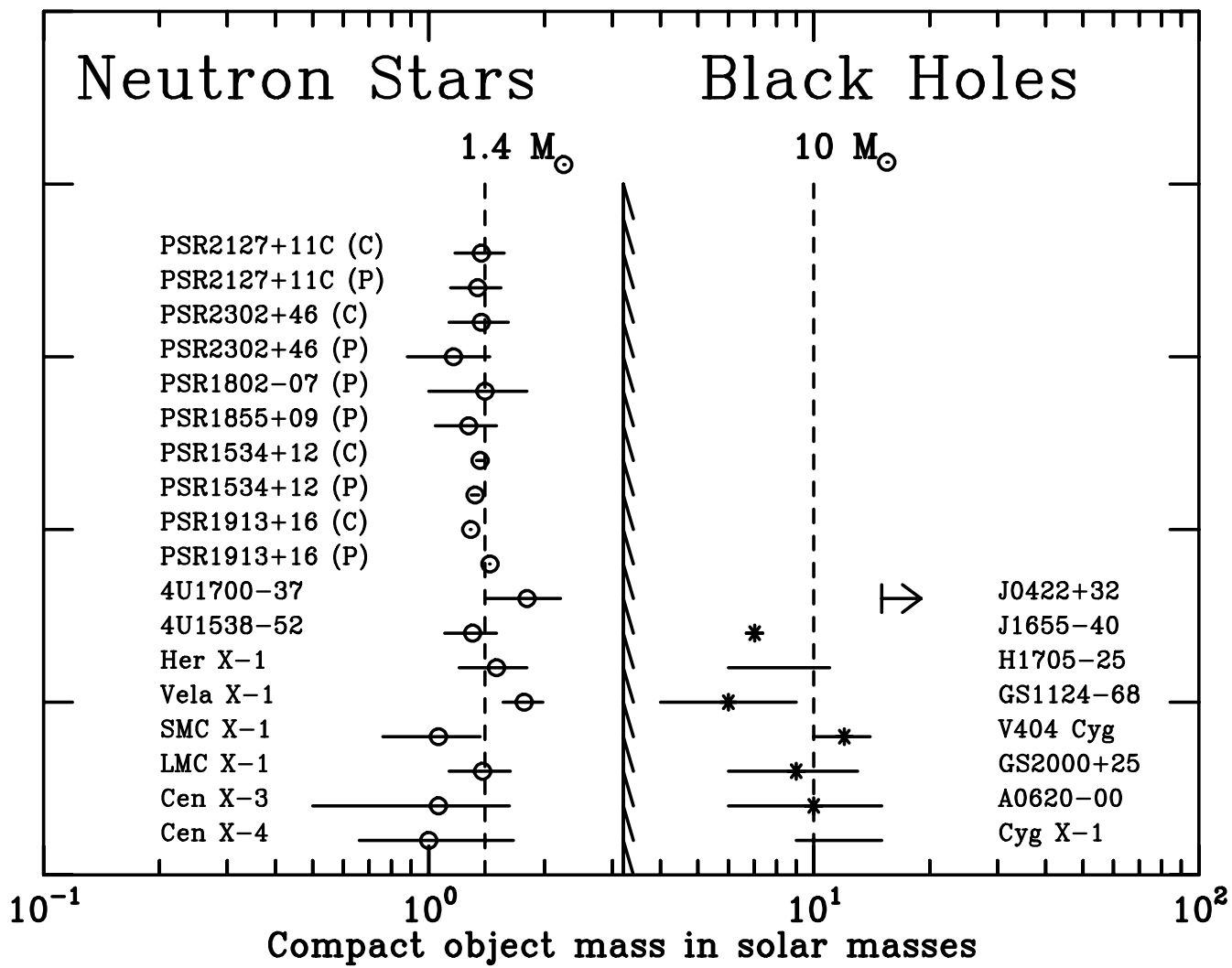
{\bf Figure 1:} 
Mass distribution of neutron stars and black holes for which masses have
been directly measured (Gies \& Bolton 1986; Nagase 1989; 
Shahbaz, Naylor \& Charles 1993; Thorsett et al. 1993; 
Shahbaz, Naylor \& Charles 1994a; Shahbaz et al. 1994b; 
van Paradijs \& McClintock 1995; Beekman et al. 1996; 
Orosz \& Bailyn 1996; Beekman et al. 1997; Remillard et al. 1996; 
Reynolds et al. 1997; Shahbaz, Naylor \& Charles 1997).
The components of the binary pulsars are marked with with P (pulsar) and 
C (companion). Also shown are the X-ray pulsars. The neutron star
systems lie at $\sim 1.4\, M_\odot$, whereas the black-hole candidates
seem to cluster around $\sim 10\, M_\odot$. The strongest black hole
candidates are V404 Cyg and J1655-40. The vertical hashed line represents
the Rhoades/Ruffini limit of $3.2\, M_\odot$. \\

\noindent
{\bf Figure 2:} 
Representative mass/radius relations for (a) Q-stars and 
(b) standard neutron stars. Note that the scaling and precise form 
of the Q-star curve depend on parameter values which 
uncertain. \\

\noindent
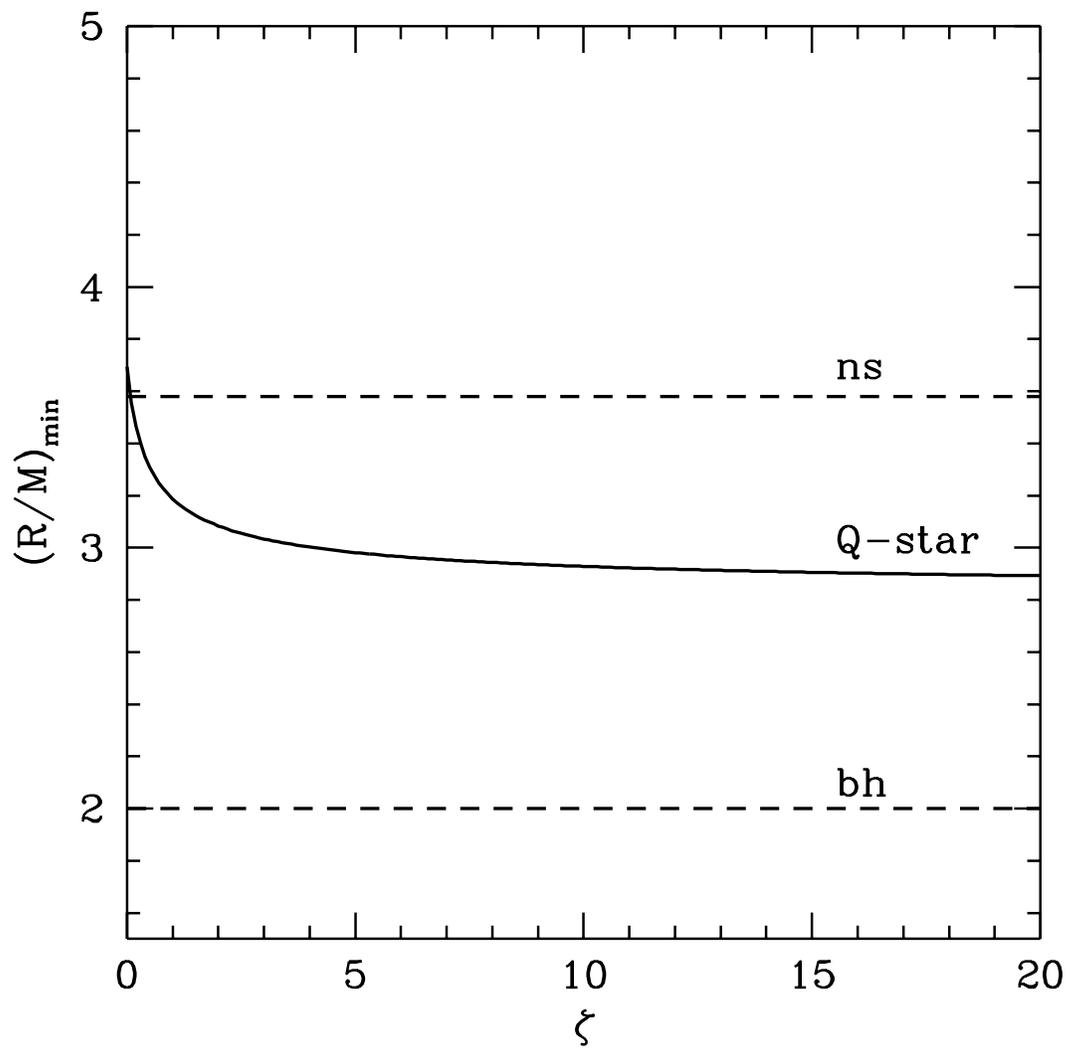
{\bf Figure 3:} 
$(R/M)_{min}$, the minimum value of $(R/M)$ for stable non-rotating
Q-star models, is plotted as a function of $\zeta$. Also shown, for
comparison, are values for a representative neutron star equation of 
state and for a black hole. For the 
Q-star, $(R/M)_{min}$ tends towards a constant value of $2.8$ as $\zeta \to
\infty$: Q-stars can be very compact.\\

\noindent
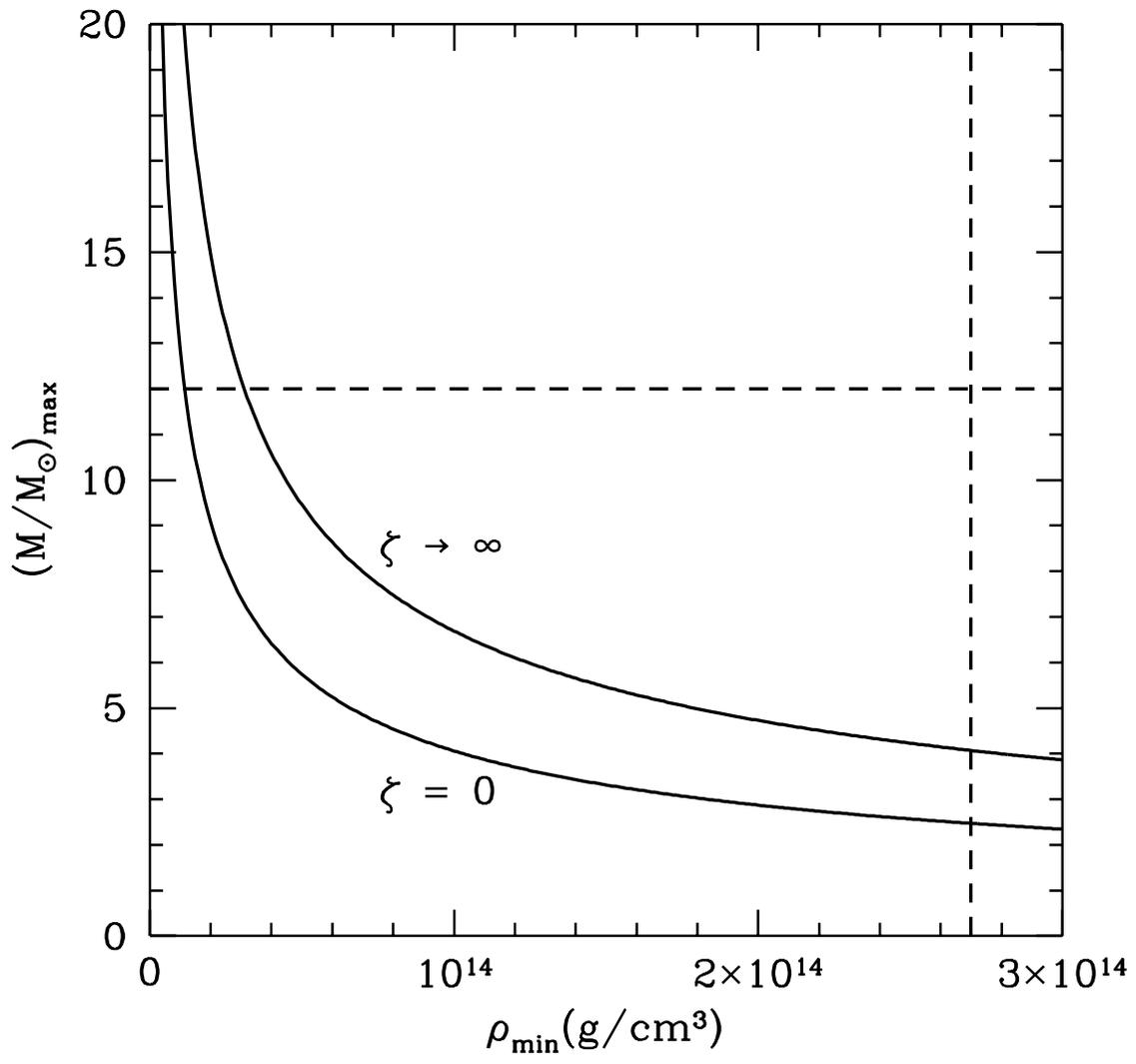
{\bf Figure 4:} 
The maximum mass of non-rotating Q-star models is plotted as a function of
the threshold density for Q-matter $\rho_{min}$. Curves are drawn for the
limiting cases $\zeta = 0$ and $\zeta \to \infty$. The horizontal
dashed line corresponds to the mass of the compact object in V404 Cyg;
the vertical dashed line marks the nuclear matter density $\rho_{nm}$. 
To allow for a Q-star with a mass as high as $12\, M_\odot$,
the threshold density would have to be about an order of magnitude below
nuclear matter density, or less. \\

\noindent
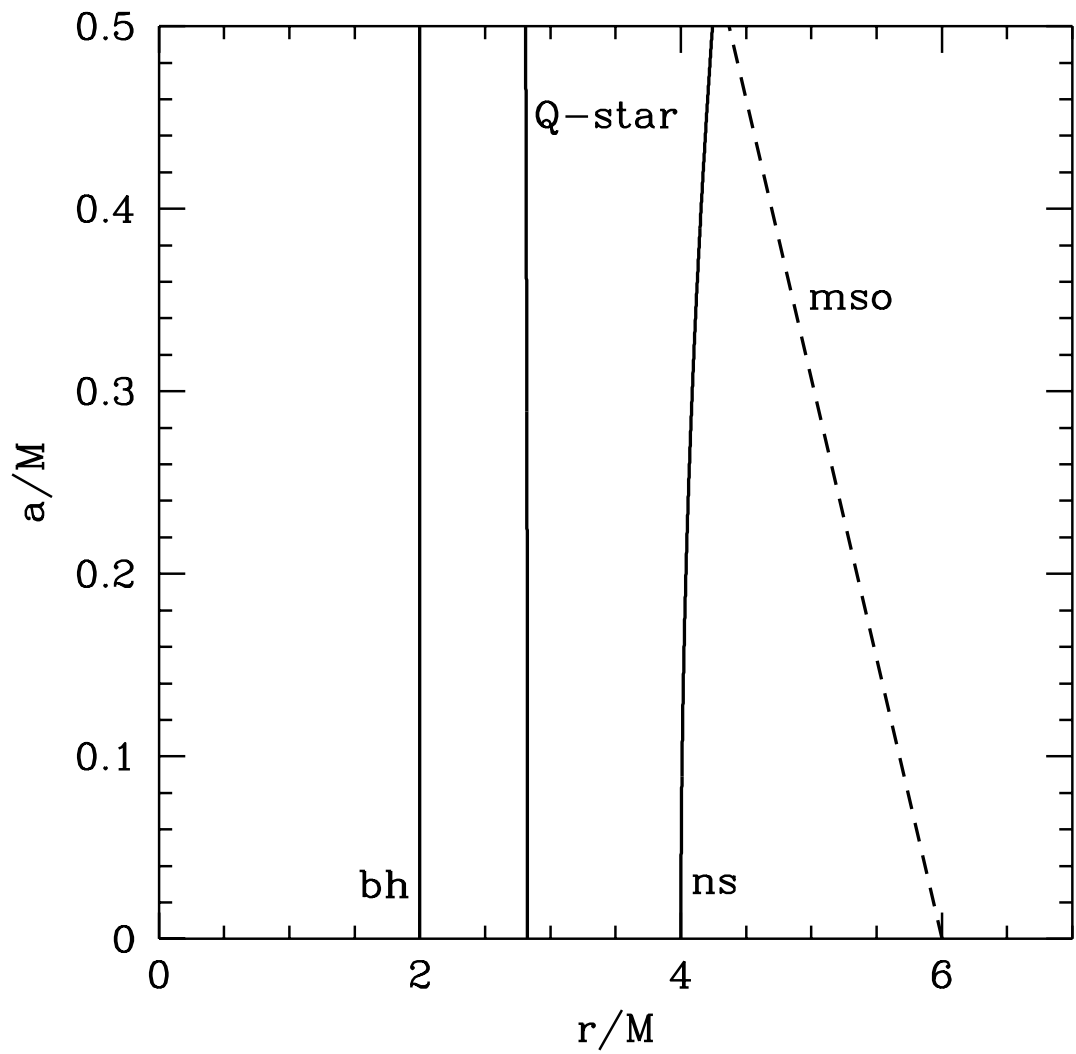
{\bf Figure 5:} 
Equatorial radii of a black hole, an extreme Q-star, a typical compact
neutron star and the location of the marginally stable orbit are plotted
as functions of $a/M$. These are (coordinate independent) circumferential 
proper-distance radii calculated consistently within the slow rotation
approximation. The equatorial radius $R_{eq}$ of the black hole
measured in this way is constant, independent of $a/M$. $R_{eq}/M$ for
the most extreme Q-stars actually {\it decreases} very slightly as $a/M$
is increased within the slow rotation regime, which is in contrast with
the behaviour of the neutron star. 

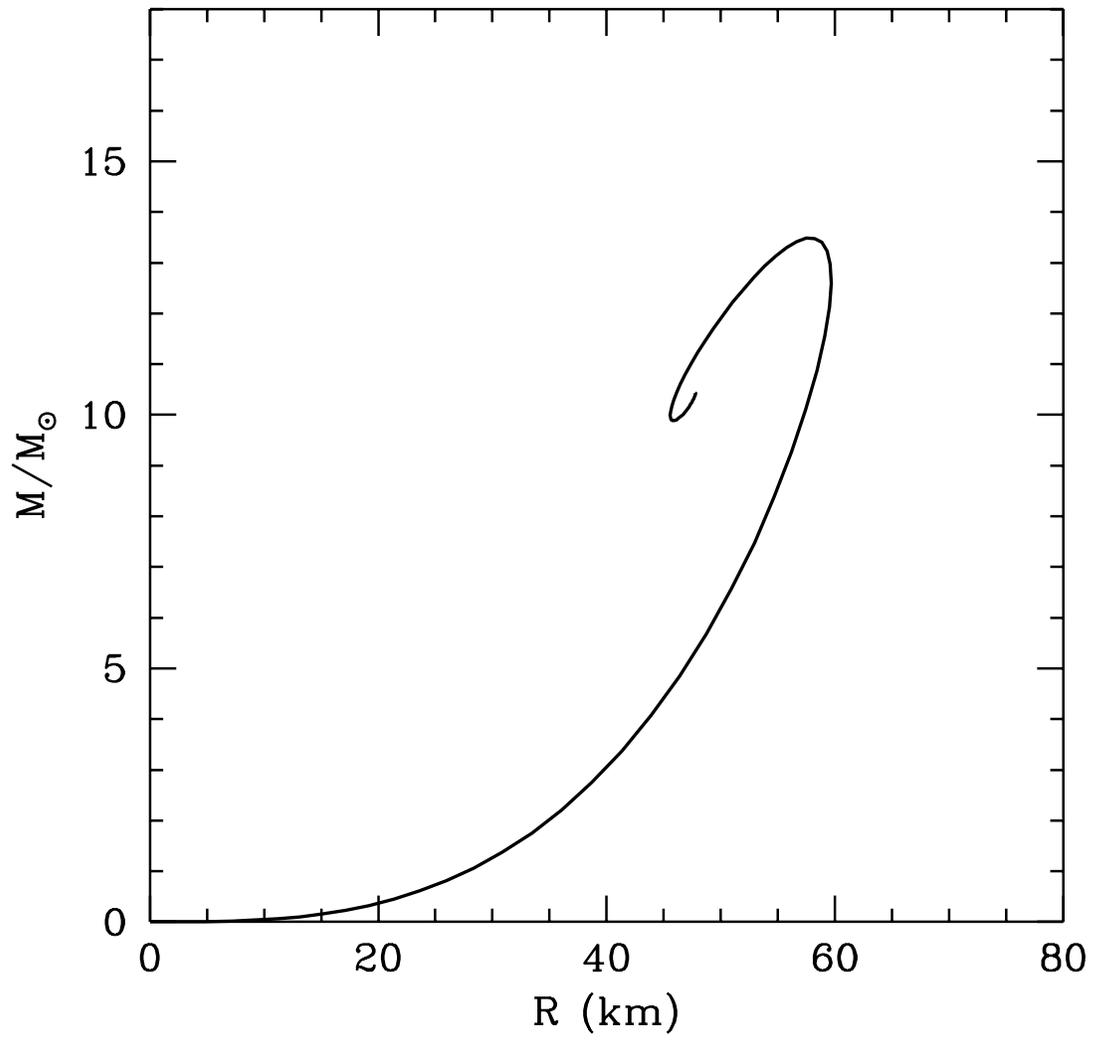
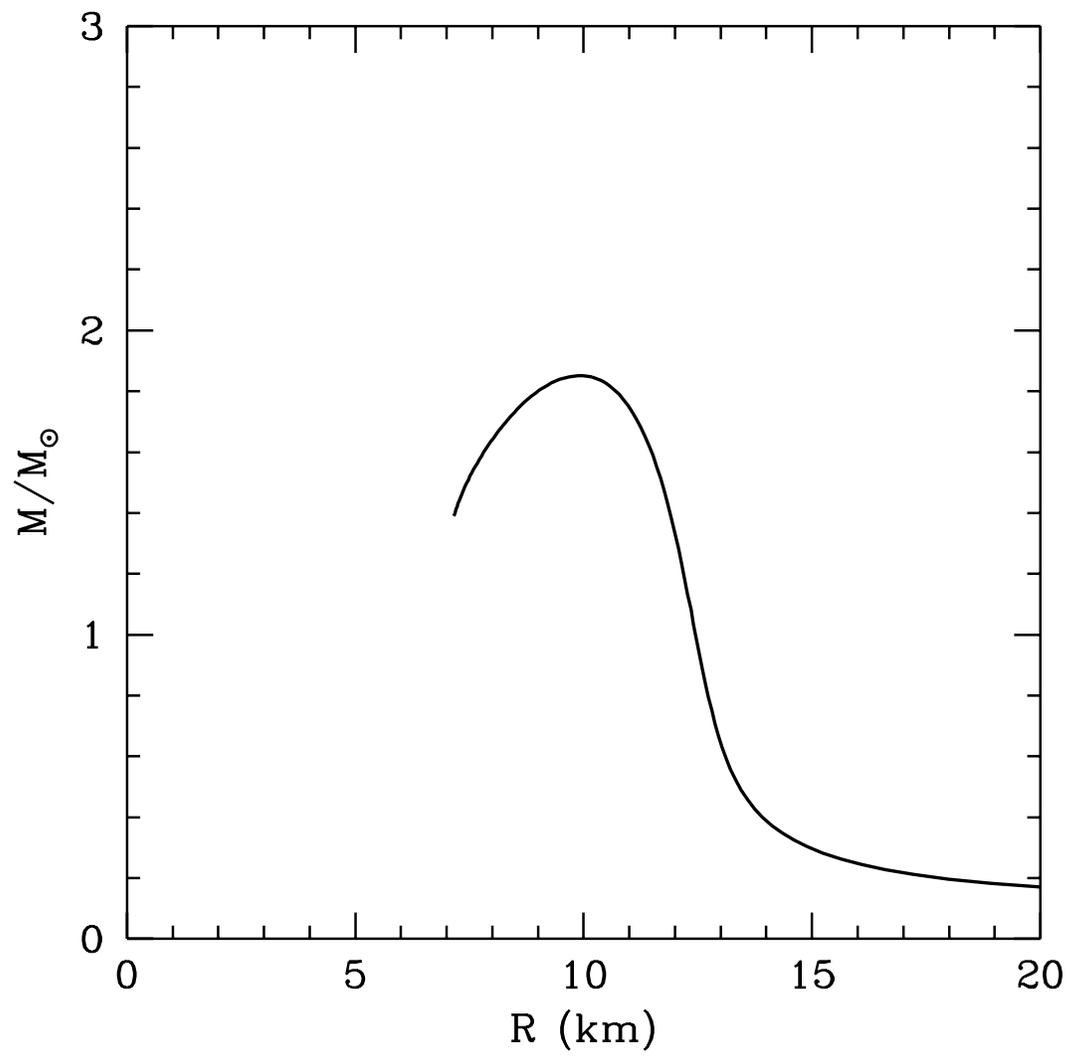
\end{document}